**Existence of a critical canting angle of magnetic moments to induce multiferroicity in the Haldane spin-chain system, Tb$_2$BaNiO$_5$**

Ram Kumar,[1] Sudhindra Rayaprol,[2] Sarita Rajput,[3] Tulika Maitra,[3] D.T. Adroja,[4,5] Kartik K Iyer,[1] Sanjay K Upadhyay,[1] and E.V. Sampathkumaran[1,*]

[1]*Tata Institute of Fundamental Research, Homi Bhabha Road, Colaba, Mumbai 400005, India*

[2]*UGC-DAE-Consortium for Scientific Research, Mumbai Centre, BARC Campus, Trombay, Mumbai – 400085, India*

[3]*Department of Physics, Indian Institute of Technology, Roorkee-247667, Uttarakhand, India*

[4]*ISIS Pulsed Neutron and Muon source, STFC Rutherford-Appleton, Laboratory, Harwell Campus, Didcot, Oxfordshire, 0X11 0QX, United Kingdom*

[5]*Highly Correlated Matter Research Group, Physics Department, University of Johannesburg, P.O. Box 524, Auckland Park 2006, South Africa*

**Abstract**

We report an unusual canted magnetism due to 3d and 4f electrons, occupying two different crystallographic sites, with its consequence to electric dipole order. This is based on neutron powder diffraction measurements on $Tb_2BaNiO_5$ (orthorhombic, *Immm* centrosymmetric space group), exhibiting Néel order below ($T_N=$) 63 K, to understand multiferroic behavior below 25 K. The magnetic structure is made up of Ni and Tb magnetic moments, which are found to be mutually canted in the entire temperature range below $T_N$, though collinearity is seen within each sublattice, as known in the past. First-principles density functional theory calculations (GCA+SO and GGA+U+SO approximations) support such a canted ground state. The intriguing finding, being reported here, is that there is a sudden increase in this Tb-Ni relative canting angle at the temperature (that is, at 25 K) at which spontaneous electric polarization sets in, with bond distance and bond angle anomalies. This finding emphasizes the need for a new spin-driven polarization mechanism – that is, a critical canting angle coupled with exchangestriction - to induce multiferroicity in magnetic insulators with canted spins.

PACS numbers: 75.85.+t; 71.15.Mb; 75.25.+z; 75.47.Lx

The study of the materials exhibiting multiferroicity – a phenomenon arising from a coupling between two ferroics orders which were historically considered to be mutually exclusive, e.g., ferroelectric and magnetic order parameters – is an active topic of research [1-14]. Over the past decade, various theoretical models have been proposed to explain such a spin-driven ferroelectricity [see, for instance, 1, 13–24], with the magnetism from transition-metal ions commonly known to trigger such a phenomenon. While models based on exchangestriction have been applied to collinear (CL) spin systems, Dzyaloshinski-Moriya interaction (DMI), which is asymmetric exchange interaction ($S_i$ x $S_j$) between spins at different sites, has been believed to govern multiferroic behavior for the materials with noncollinear (NCL) magnetic structures (that is, for spatially rotating magnetic structure). Conventional DMI-based models, such as Katsura, Nagaosa, and Balatsky (KNB) model [Ref. 16], are applicable to those in which normal vector of the spin spiral plane is *perpendicular* to the propagation vector (as in cycloidal magnetic structure) and does not involve lattice degree of freedom. The DMI interaction via a third ligand ion was also proposed by Sergienko and Dagotto [17] to lead to electric polarization due to displacement of the ligand in transition metal systems in which superexchange mechanism mediates magnetic ordering. However, such DMI-based models could not explain the origin of ferroelectricity in proper-screw type magnetic systems - as in delafossites (e.g., $CuCrO_2$) [18] - in which normal vector of the spins at two different sites is *parallel* to the propagation vector. Among various other theories to describe canted-spin-caused ferroelectricity [19-24], a 'local' approach by Kaplan and Mahanti [19] offered an explanation, based on the formation of the dipole moment caused just by a pair of canted spins without the need to invoke magnetic structures like that of spiral. Thus, this theory is a more general form of DMI explaining multiferroicity, adding an additional contribution to KNB coupling. Subsequently, Miyahara and Furukawa [24] proposed a microscopic model for such a 'local' spin-pair-dependent electric polarization. In a nutshell, these theories [23, 24] demand that multiferroicity can be observed in any canted antiferromagnet, owing to non-KNB coupling, apparently not restricted to d-metal ions. Clearly, NCL magnetic structures can present surprising situations in the field of magnetism [25]. Therefore, there is a need to search for materials for new anomalies involving canted-spins in general to advance the knowledge in the field of multiferroics. This is the motivation of the present work.

In this article, we provide an evidence in a globally centrosymmetric material, for a new spin-driven polarization mechanism. The conclusion is based on the neutron powder diffraction (NPD) studies on $Tb_2BaNiO_5$ (space group: *Immm*, see Ref. 27 for crystal structure*),* which was shown to be an 'exotic' multiferroic below 25 K recently [26]. Density functional theoretical (DFT) calculations were carried out to support canted magnetic ground state, derived from NPD.

The compound under investigation is a derivative of a spin-chain compound, $Y_2BaNiO_5$ [28], in which Y is non-magnetic. The spin-chain, running along a-axis (inset, Fig. 1), is made up of integer spins of Ni and hence this Y compound is characterized by a gap between non-magnetic ground state and excited state (known as Haldane-gap). The interchain interaction is weak. In sharp contrast to this, when Y is replaced by magnetic-moment containing rare-earths (R), antiferromagnetic (AFM) ordering gets triggered at both R and Ni sites, interestingly, at the same temperature in the family $R_2BaNiO_5$. These compounds with a moment on R are spin-driven multiferroics [26, 29-31]. The magnetic structure is characterized [32, 33] by the temperature-independent propagation vector k = (1/2, 0, 1/2). Exotic nature of the Tb compound is due to the following: The Néel temperature ($T_N$= 63 K, also called $T_{N1}$ here) is the highest within this series and the observed value of magnetodielectric coupling (18%) is the largest ever reported for a polycrystalline compound, attributable to single-ion 4f orbital anisotropy [26]. In addition, this compound exhibits another subtle, but distinct, magnetic anomaly at ($T_{N2}$=) 25 K which induces electric dipole ordering below this temperature only (and hence $T_{N2}$ is called ferroelectric Curie temperature, $T_C$). Since no information about canting angle behavior across 25 K was presented in Ref. 32, we considered it absolutely essential to reinvestigate this compound by NPD carefully and to augment it with DFT calculations to throw light on the origin anomalies of this compound.

NPD experiments were carried out on polycrystalline sample of $Tb_2BaNiO_5$ on WISH diffractometer on target station (TS-2) of the Rutherford Appleton Laboratory, UK. NPD patterns were obtained from 2 K to 80 K in steps of 2 K. FullProf [34] program was used to refine the nuclear and magnetic structures using the data measured in detector banks at average scattering angles (2Θ) of 27º, 58º, 90º, 122º and 153º each covering 32º of the scattering plane. Electronic structure and magnetic ground state of $Tb_2BaNiO_5$ are obtained using projector augmented plane-wave (PAW) based method within the DFT framework as implemented in the Vienna Ab-initio Simulation Program (VASP) [35]. The details of methodology used in our calculations are given in the Supplementary Materials [27].

The NPD pattern collected at $T = 80$ K was refined successfully for the orthorhombic structure in the space group *Immm,* in good agreement with the previous reports [28, 31]. The details of the structural parameters are given in Table 1 of Supplemental Material [27]. In Fig. 1, the observed ($I_{obs}$) NPD patterns along with refined profile ($I_{cal}$) are shown for $T = 2, 60$ and 80 K, which indicate good agreement. As expected, below $T_N$ (say, at ~60 K), additional peaks could be seen due to magnetic ordering, as known earlier [32, 33]. These magnetic Bragg peaks could be indexed by invoking the propagation vector, $k = (½, 0, ½)$. The magnetic structure at 2 K resembles the model reported in the literature [32]. In fact, the magnetic structure of $Ho_2BaNiO_5$ was taken as the starting model for refinement [36]. As the temperature is lowered below 63 K, the magnetic Bragg peaks gain intensity due to the ordered moments of Tb and Ni. The magnetic structures at 2, 26 and 60 K are shown in figure 2. In the inset of Fig. 3a, the variation in the magnitude of θ as a function of temperature is plotted for Ni and Tb magnetic-moments. The angle here refers to the magnitude of the angle between the moment-vector and the crystallographic (positive direction) *c*-axis. The magnetic-moment of Ni is canted in the entire *T*-range below $T_{N1}$, with the moment orienting almost along *c*-axis at the onset of magnetic order. But /θ/ increases with a gradual lowering of temperature, attaining a value of about 45° at 4.2 K. Thus, there is a large variation of θ of Ni moment below $T_{N1}$. However, the Tb moment is oriented close to *c*-axis (θ= 1 to 7°) at all temperatures with a weak *T*-dependence below $T_{N1}$. But, what is intriguing is that there is a sharp increase in the canted angle subtended by Ni with the *c*-axis at 25 K from about 18° in a narrow temperature interval reaching a saturation value below about 15 K. Naturally, the difference (Δθ) in the canting angle of nearby Ni and Tb moments (oriented towards, say, positive *c*-direction) increases sharply below this temperature (Fig. 3a). There is a noticeable change in Ni/Tb magnetic moments (Fig. S2, Ref. 27). It is worth noting that the shape of the plot of /Δθ/ (Fig. 3a) resembles the plot of electric polarization versus temperature, reported earlier (see Fig. 4b in Ref. 26), thereby establishing a close correlation between these two.

It is to be noted that there are sudden changes at 25 K in the angles of O1-Ni-O1 and Ni-O1-Tb as shown in figures 3b and 3c. O1-Ni-O1 bond angle remains steady around 78.905º up to 40 K, and, below 40 K, it starts increasing gradually peaking at 24 K. Below 24 K, it starts decreasing to around 78.895º. The overall change in the absolute values is small, but it is much bigger than the experimental errors. The Ni-O1-Tb bond angle also undergoes a change, with this angle increasing below 24 K.

We observed that the influence of coupled ordering at 25 K is felt in the lattice constants as well (Fig. 4a), determined from NPD patterns. The lattice constant *a* increases, and *b* and *c* decrease with decreasing temperature, but exhibit sudden changes around 25 K. It may be noted that there is a gradual change of slope even near $T_{N1}$. Since it is more instructive to see how bond lengths are getting influenced at this temperature, we have plotted *T*-dependence of bond lengths for Ni-O1 and Ni-O2 (lying along the chain, called "apical" distance in Ref. 32) in figures 4b and 4c. We can confidently state that the Ni-O1 bond length tends to show an upward trend below 25 K, with respect to the positive temperature coefficient seen above 25 K. There is of course an influence on Ni-O2 distances as well, in the sense that the slope value changes gradually around this temperature. A careful look at the analyzed data revealed that the z value for Tb with the coordinate (1/2, 0, z) exhibits a weak, but observable, change in the vicinity of $T_{N1}$ as well as at $T_{N2}$, suggesting that Tb displacement possibly causes local electric polarization. On the basis of displacement, we infer that the local polarization along b-direction is favored, than along c-direction.

In short, the NPD data show that there is an apparent critical (*relative*) canting angle of Ni and Tb magnetic moments that has to be exceeded for electric dipole ordering to get triggered by magnetic order – which is a fascinating observation. As shown above, there is a weak, but a sudden, change in the bond distances and the bond angles subtended by magnetic moments with oxygen at 25 K. *These are the key findings of this experimental work, which was not reported in earlier NPD investigations [32].* Note that the sub-lattice magnetic structure is of a collinear type for both Ni and Tb. In other words, there is no evidence for cycloidal magnetic structure, even when one enters multiferroic region (that is, below 25 K), thereby raising a question on the applicability of conventional DMI based models [16] to describe multiferroicity in this case. Even at $T_{N1}$, there is a lattice strain (possibly due to symmetric exchange from the intra-sublattice collinearity), but it is still weaker, as inferred from a weak slope change in the plots of lattice constants versus temperature (Fig. 4), but the strain below 25 K only results in electric dipole ordering. This finding establishes that it is the large difference in local canting angles of Ni and Tb moments below 25 K that apparently results in a relatively enhanced lattice strain leading to net spontaneous electric polarization. These observations support the need for an exchangestriction-based model even to NCL magnetic insulators, also when a rare-earth with well-localized magnetic moments (not favorable to super-exchange interaction) is involved in interatomic canting to trigger multiferroicity. The 'local spin-canting' theory of Ref. 19 appears to be more general in this respect (that is, without insistence of any spiral magnetic structure) and it is therefore of interest to extend this model to take into account exchangestriction. [In the range 25 – 63 K, possible local electric dipoles created by canting appears to average out over a large volume, making the net spontaneous polarization zero]. At this juncture, it is worth stating that our initial studies on $Tb_{2-x}Y_xBaNiO_5$, investigated up to x= 1.5, suggest that $T_{N2}$ and $T_C$ get reduced linearly with x (Fig. S3 in Ref. 27), almost scaling with the concentration of Tb. This finding favors local canting involving Tb 4f (consistent with the role of Tb, inferred from z-coordinate values as well above). Finally, we admit that we are not able to resolve noncentrosymmetry at the onset of electric dipole order from the present NPD results. We reconcile this by the fact that the lattice distortions are so small that the resulting loss of inversion symmetry escapes detection in neutron and x-ray diffraction, as known for some other multiferroics, even in the recent past [37, 38]. We are not however handicapped by this assumption to draw the present conclusions.

In order to render support to the magnetic state obtained from NPD, we carried out first-principles DFT calculations. The total energy calculations were performed using the experimental structural parameters for various magnetic configurations [Fig. S4 in Ref. 27] such as ferromagnetic (FM), ferrimagnetic, and different AFM orders including the one observed in experiments within GGA, GGA+U and GGA+U+SO approximations. We noted that, within GGA+SO approximation, the electronic structure is metallic which is not consistent with experimentally observed insulating behavior of this spin-chain family. Therefore, a finite Coulomb correlation energy, U, is essential in the calculations. We further computed total energies assuming collinear and non-collinear spin configurations (with respect to c-axis) for the AFM order observed experimentally (AFM-CL and AFM-NCL respectively) within GGA+U+SO approximation. For CL magnetic state, the magnetic moments of both Tb and Ni ions were fixed parallel to the c-axis. However, for the case of NCL magnetic state, the magnetic moments of Ni ions were allowed to rotate in the a-c plane; the Tb moments were fixed along c-axis as neutron measurements show very small θ for Tb and the end results are not affected by such marginal changes. In the NCL case, the initial magnetic state with Ni moments canted at 45° with respect to c-axis, was allowed to relax over a self-consistency cycle until it reached the energy minimum. The value of $U_{eff}$= U – J (where U is the Coulomb correlation and J is Hund's exchange) was varied systematically between 0 and 4.5 eV for Ni, while for Tb the values were fixed at 0 and at a value typical of heavy rare-earths, e.g., 7eV. Comparing total energies of various magnetic states including the CL and NCL magnetic states, we observe that within GGA+SO approximation, the ferromagnetic state is the ground state. On incorporating U, the ground state becomes non-collinear AFM which is consistent with that found in the neutron diffraction measurements described above. Further, when we vary $U_{eff}$ for Ni d-states (keeping $U_{eff}$ fixed at 7eV for Tb f-states) up to a critical value of $U_{eff}$ for Ni (which is around 2eV here), the ground state remains non-collinear. However, on increasing $U_{eff}$ beyond 2eV, the CL magnetic state becomes the lowest energy state (see Table II in Supplemental Material). From our NCL calculations, we have computed θ of Ni moment (Table III and Fig. S5 in Supplemental Material) which are seen to increase with U applied to Ni and, for $U_{eff} \geq 1.5$ eV, it lies in the range of 30°-40° which are quite close to experimentally observed value at 2 K. Finally, we have fully optimized the experimental crystal structure for FM and AFM-CL configurations. Optimized structural parameters are listed in Table IV of Supplemental Material. We can clearly see that in the case of AFM-CL, Ni-O-Ni bond angle along the chain (i.e. a-axis) is not exactly 180º as found in the case of FM, establishing local distortion.

In conclusion, the article brings out a situation in which a critical canting angle of a pair of spins (Ni 3d and Tb 4f) determines the onset of magnetoelectric coupling. It is the existence of two different canted regime below Néel order in a single compound that enables us to pose a question whether the concept of a critical canting angle for such an adjacent pair of magnetic ions with favorable exchangestriction, is relevant for multiferroicity. Additionally, it is worth noting that one of the pairs involved in canting contains well-localized 4f orbital with non-zero orbital angular momentum, occupying a site different from the d-ion in the crystal structure, unlike in many other spin-driven multiferroics in which (spatially extended) d-orbitals occupying the same site cause multiferroicity. Dong et al [39] emphasized recently the need to identify multiferroics in which f-moments play a key role to understand the role of spin-orbit coupling to mediate electric polarization. This Tb-based material could be a case to address this issue. In short, this work is a step forward in canted-spin-caused multiferroicity.

One of us (E.V.S) would like to thank D. Khalyavin for his help during neutron diffraction measurements, Ganapathy Vaitheeswaran for preliminary DFT calculations on simple ferromagnetic and antiferromagnetic structures, and K. Maiti for his comments on the manuscript. We thank ISIS facility for providing beam time on WISH, RB1810006.

**References**
*Corresponding author e-mail address: sampathev@gmail.com
[1]. D. I. Khomskii, Physics **2**, 20 (2009); J. van den Brink and D.I. Khomskii, J. Phys. Condens. Matter, 20, 434217 (2008).
[2]. N. A. Spaldin, S. W. Cheong, and R. Ramesh, Physics Today **63**, 38 (2010).
[3]. M. Fiebig, T. Lottermoser, D. Meier, M. Trassin; Nature Reviews Materials **1,** 16046 (2016).
[4]. S. Dong, J.-M. Liu, S.-W. Cheong, and Z. Ren, Adv. Phys. **64**, 519 (2015)
[5]. M. Fiebig, J. Phys. D: Appl. Phys. **38,** R123 (2005).
[6]. G. A. Smolenskii, I. Chupis, Sov. Phys. Usp. **25**, 475 (1982).
[7]. A. Filippetti, N. A. Hill, J. Magn. Magn. Mater. **236**, 176 (2001).
[8]. O Heyer, N Hollmann, I Klassen, S Jodlauk, L Bohaty, P Becker, J A Mydosh, T Lorenz and D Khomskii, J. Phys.: Condens. Matter **18**, 39 (2006).
[9]. N. Hur, S. Park, P.A. Sharma, J.S. Ahn, S. Guha, and S.-W. Cheong, Nature **429,** 392 (2004).
[10]. T. Kimura, T. Goto, H. Shintani, K. Ishiazaka, T. Arima, and Y. Tokura, Nature (London) **426,** 55 (2003).
[11]. G. Lawes, A. B. Harris, T. Kimura, N. Rogado, R. J. Cava, A. Aharony, O. Entin-Wohlman, T. Yildrim, M. Kenzelmann, C. Broholm, and A. P. Ramirez, Phys. Rev. Lett. **95**, 087205 (2005).
[12]. T. Kimura, J. C. Lashley, and A. P. Ramirez: Phys. Rev. B **73** 220401(R) (2006).
[13]. Y. J. Choi, H. T. Yi, S. Lee, Q. Huang, V. Kiryukhin, and S.-W.Cheong: Phys. Rev. Lett. **100,** 047601 (2008).
[14]. Y. Naito, K. Sato, Y. Yasui, Y. Kobayashi, Y. Kobayashi, and M. Sato: J. Phys. Soc. Jpn. **76**, 023708 (2007).
[15]. M. Mostovoy, Phys. Rev. Lett. **96**, 067601 (2006).
[16]. H. Katsura, N. Nagaosa, and A. V. Balatsky, Phys. Rev. Lett. **95**,057205 (2005).
[17]. I. A. Sergienko and E. Dagotto, Phys. Rev. B **73,** 094434 (2006).
[18]. N. Terada, J.Phys.: Condens. Matter **26,** 453202 (2014).
[19]. T. A. Kaplan and S. D. Mahanti, Phys. Rev. B **83**, 174432 (2011).
[20]. J. Hu, Phys. Rev. Lett. **100**, 077202 (2008).
[21]. A. B. Harris, T. Yildirim, A. Aharony, and O. Entin-Wohlman, Phys. Rev. B **73,** 184433 (2006).
[22]. T. Arima, J. Phys. Soc. Jpn. **76,** 073702 (2007).
[23]. C. Jia, S. Onoda, N. Nagaosa, and J. H. Han, Phys. Rev. B **76,** 144424 (2007).
[24]. S. Miyahara and N. Furukawa, Phys. Rev. B **93,** 014445 (2016).
[25]. E. Bousquet and Andres Cano, J. Phys.: Condens Matter **28,** 123001 (2016).
[26]. S. K. Upadhyay, P.L. Paulose and E.V. Sampathkumaran, Phys. Rev. B **96,** 014418 (2017).
[27]. See, Supplemental Material
[28]. J. Darriet and L. P. Regnault, Solid State Commun. **86,** 409(1993).

[29]. T. Basu, V. V. Ravi Kishore, S.Gohil, K. Singh, N. Mohapatra, S. Bhattacharjee, B. Gonde, N. P. Lalla, P. Mahadevan, S. Ghosh, and E. V. Sampathkumaran, Sci. Rep. **4,** 5636 (2014).
[30]. S. Chowki, Tathamay Basu, K. Singh, N. Mohapatra, and E. V. Sampathkumaran**,** J. Appl. Phys. **114,** 214107 (2014).
[31]. Kiran Singh, Tathamay Basu, S. Chowki, N. Mohapotra, K. Iyer, P. L. Paulose and E. V. Sampathkumaran**,** Phys. Rev. B **88,** 094438 (2013).
[32]. E. Garcia-Matres, J. L. Martinez, and J. Rodriguez Carvajal, Eur. Phys. J. B **24**, 59 (2001).
[33]. E. Garcia-Matres, J.L. Garcia-Munoz, J.L. Martinez, and J. Rodriquez Carvajal, J. Magn. Magn. Mater. **149,** 363 (1995).
[34]. J. Rodriguez Carvajal, Physica B **193**, 55 (1993).
[35]. Georg Kresse, JrgenFurthmller, Phys. Rev. B. **54,** 11169 (1996).
[36]. G. Nenert and T.T.M. Palstra, Phys. Rev. B **76**, 024415 (2007).
[37]. Q. Zhang, K. Singh, F. Guillou, C. Simon, Y. Breard, V. Caignaert, and V. Hardy, Phys. Rev. B **85,** 054405 (2012).

[38]. A. Gauzzi, F. Milton, V. Pascotto Gastaldo, M. Verseils, A. Gualdi, D. Dreifus, Y. Klein, D. Garcia, A.J.A de Oliveira, P. Bordet, and E. Gilioli, arXiv1811.07182, and references therein.

[39]. S. Dong, H. Xiang, and E.Dagotto, arXiV:condmat1902.01532.

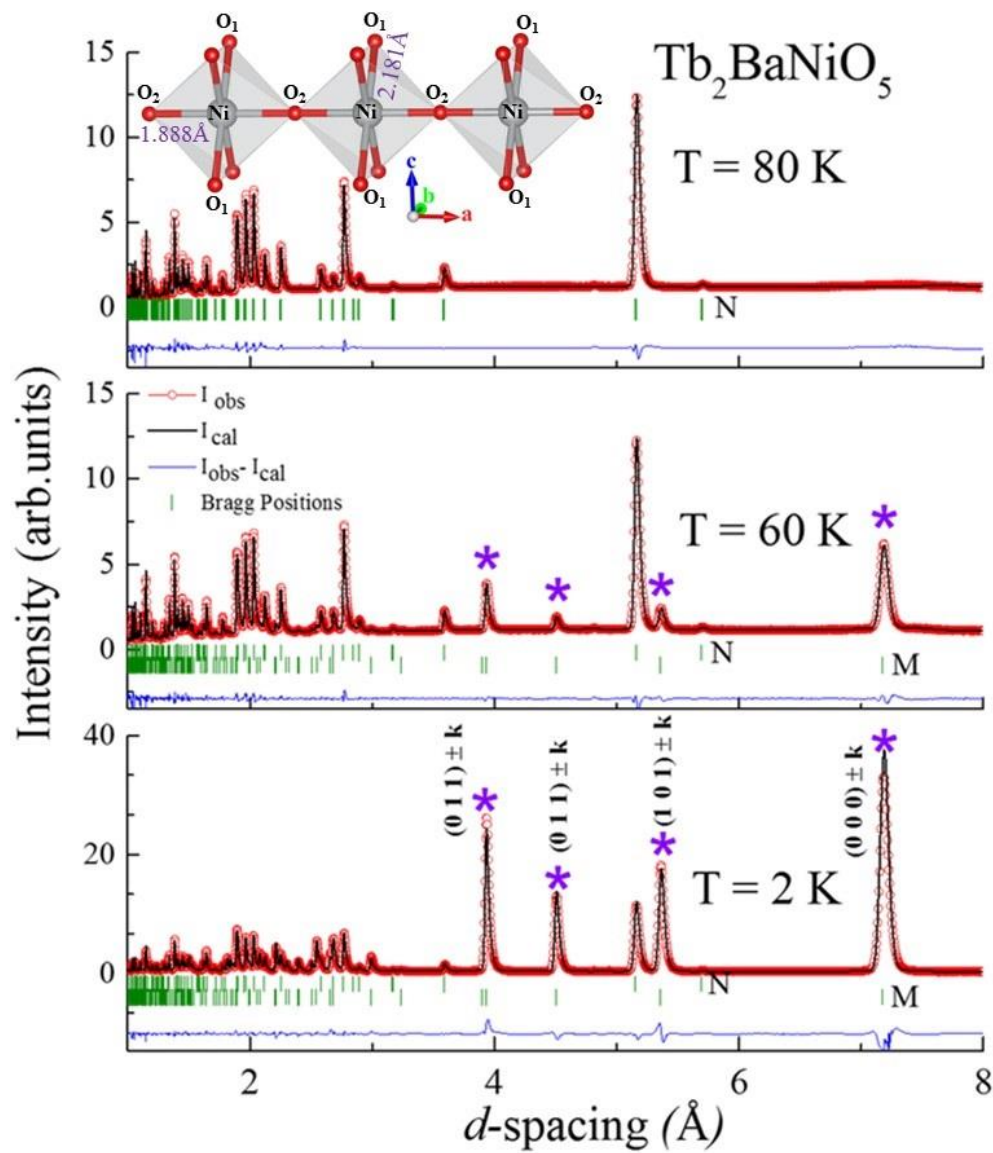

Fig: 1. Rietveld refinement of neutron diffraction patterns of $Tb_2BaNiO_5$ at selected temperatures. Inset shows the $NiO_6$ chains running along *a*-axis. In the bottom most panel, the (hkl) values for magnetic Bragg peaks where k =(1/2,0,1/2) are shown. Asterisks mark magnetic peaks. Nuclear (N) and magnetic (M) peak positions are shown by vertical green ticks.

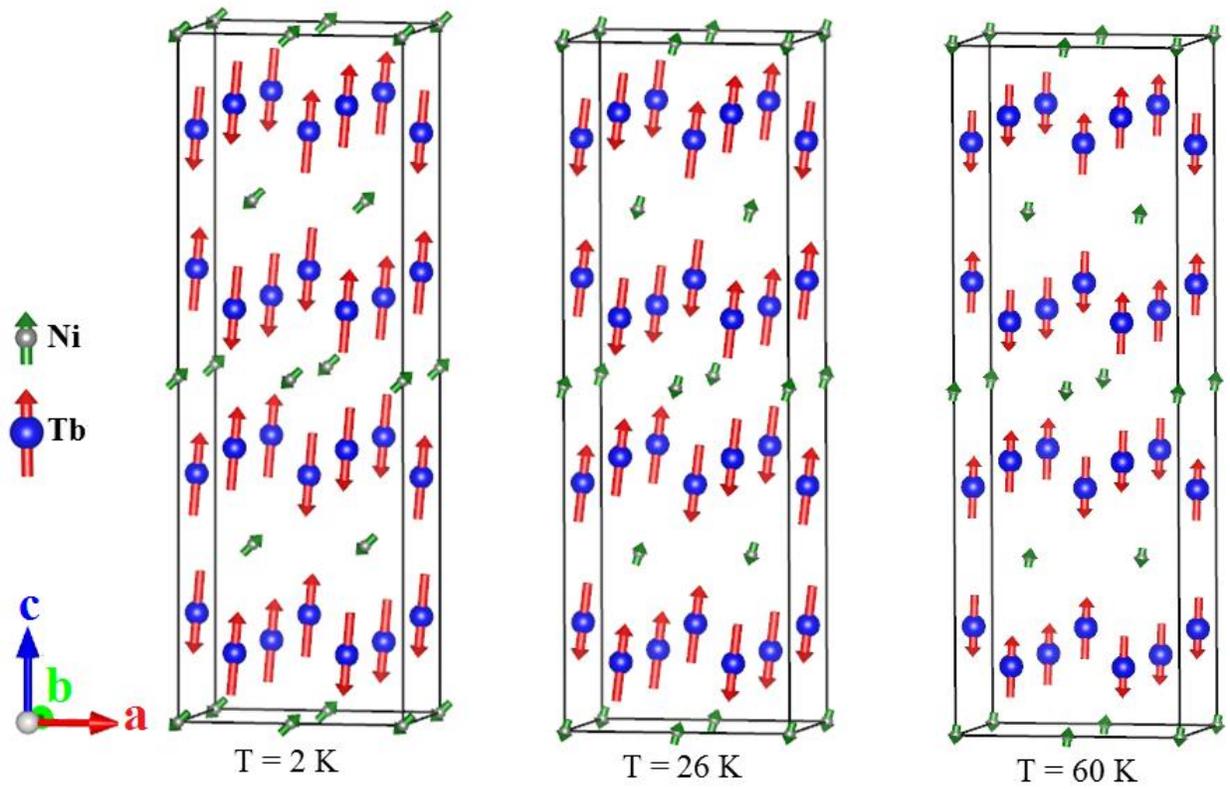

Fig.2: Magnetic structure of $Tb_2BaNiO_5$ at selected temperatures. A 2x1x2 supercell is shown here.

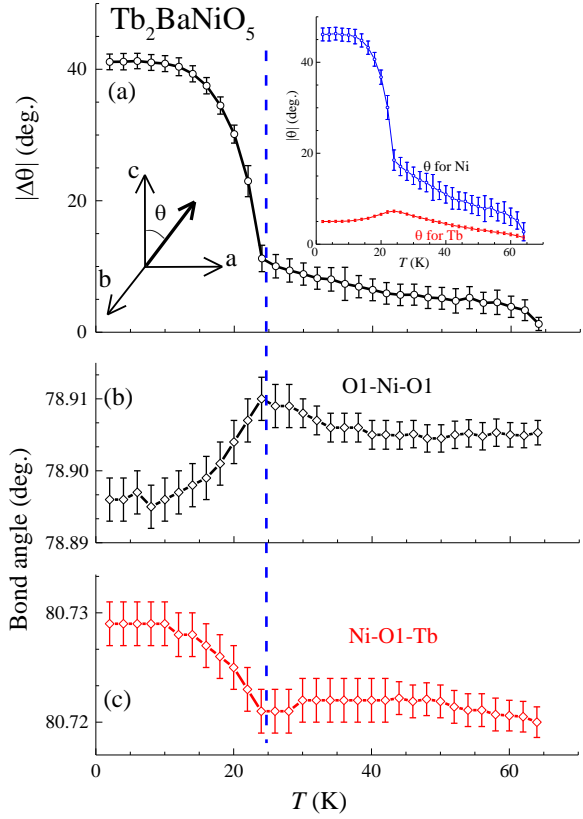

**Fig. 3: (a)** Temperature dependence of relative canting angles of Ni and Tb in Tb$_2$BaNiO$_5$ obtained from NPD data. In the inset, the magnitude of canting angles for a pair of Ni and Tb moments towards positive *c*-axis plotted. In **(b)** and **(c)**, temperature dependence of bond angles are shown. The lines through the data points serve as guides to the eyes and a vertical dashed line is shown where sudden changes occur.

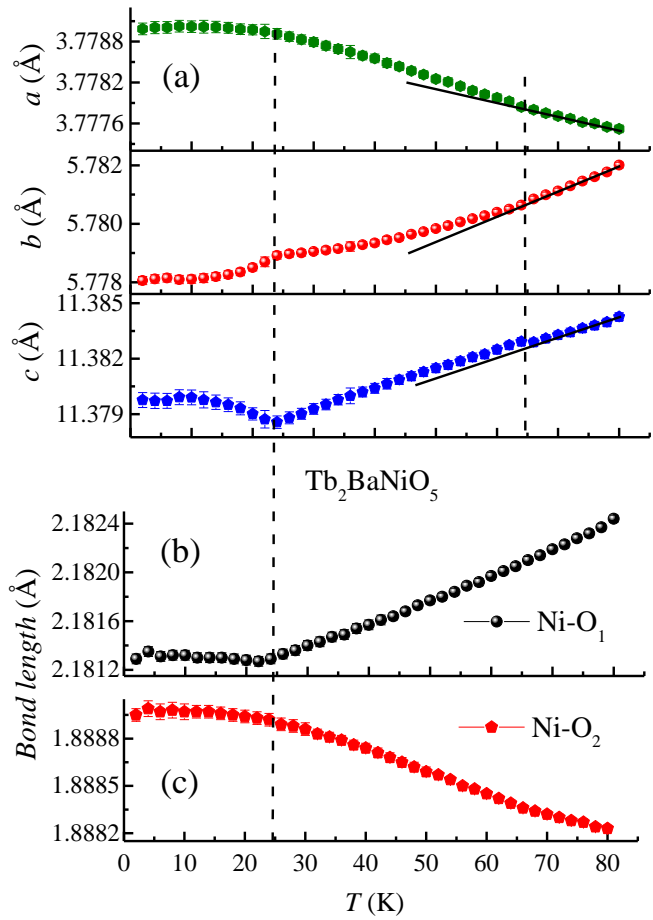

**Fig. 4**: (a) Temperature dependent lattice parameters of $Tb_2BaNiO_5$ obtained from NPD data. Vertical dashed lines are drawn to show the temperatures at which magnetic features are seen. A dashed line is drawn through the data points above $T_N$ to show that there is a subtle, but gradual increase of, lattice strain with decreasing temperature (as one enters antiferromagnetic state). In (b) and (c), bond lengths are plotted as a function of temperature.

# Supplemental Material

## Existence of a critical canting angle of magnetic moments to induce multiferroicity in the Haldane spin-chain system, $Tb_2BaNiO_5$

**Ram Kumar,[1] Sudhindra Rayaprol,[2] Sarita Rajput,[3] Tulika Maitra,[3] D.T. Adroja,[4,5] Kartik K Iyer,[1] Sanjay K Upadhyay,[1] and E.V. Sampathkumaran[1]**

[1]*Tata Institute of Fundamental Research, Homi Bhabha Road, Colaba, Mumbai 400005, India*

[2]*UGC-DAE-Consortium for Scientific Research, Mumbai Centre, BARC Campus, Trombay, Mumbai – 400085, India*

[3]*Department of Physics, Indian Institute of Technology, Roorkee-247667, Uttarakhand, India*

[4]*ISIS Pulsed Neutron and Muon source, STFC Rutherford-Appleton, Laboratory, Harwell Campus, Didcot, Oxfordshire, 0X11 0QX, United Kingdom*

[5]*Highly Correlated Matter Research Group, Physics Department, University of Johannesburg, P.O. Box 524, Auckland Park 2006, South Africa*

Here we show the crystal structure of the title compound, temperature dependence of magnetic moment values derived from neutron diffraction data, various magnetic structures considered in DFT calculations, and the dependence of canting angle of Ni moments in AFM-NCL magnetic structure on $U_{eff}$ applied to Ni ions keeping $U_{eff}$ for Tb fixed at 7eV in the figures. Structural parameters obtained by Rietveld fitting of a neutron diffraction pattern at selected temperatures, and various parameters obtained in the calculation for different input parameters are shown in the Tables.

We also show how Néel temperature and multiferroic onset temperature vary with the silution of Tb sublattice with Y. Experimental results in detail based on magnetization, heat-capacity, dielectric constant and bias current as a function of temperature and magnetic-field will be published elsewhere [1].

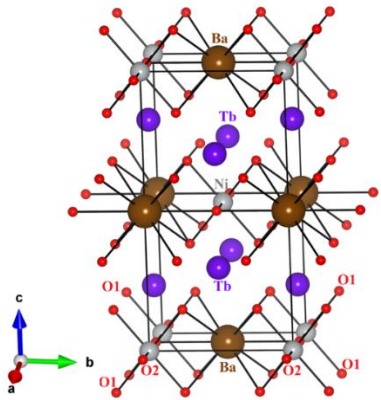

Fig: S1. Crystal structure of $Tb_2BaNiO_5$.

**Refinement methodology**: We have refined the different parameters with **a** Convolution pseudo-Voigt with back-to-back exponential functions for peak shape profile, for example, scale, cell parameters, z-position of Tb and O2 and y-position of O1, and thermal parameter ($B_{iso}$). Further, for patterns below Néel temperature, as mentioned in the main text, we have used the $Ho_2BaNiO_5$ model as in Ref. 32, refining for the magnetic moments of Tb and Ni as well.

In order to refine the magnetic structure, we have taken magnetic moments in spherical coordinates, which allows refinement of the angle of the moments with respect to crystallographic axis, in addition to the magnetic moment values.

In Fig. 1, the difference spectra are shown to bring out that the fit is good. The R-values are in the comparable range for all temperatures: R-Bragg = 3.3-5.4, $R_f$-factor= 5.4-6.1, and R-magnetic= 1.72-8.4, endorsing reliable comparison of results at different temperatures.

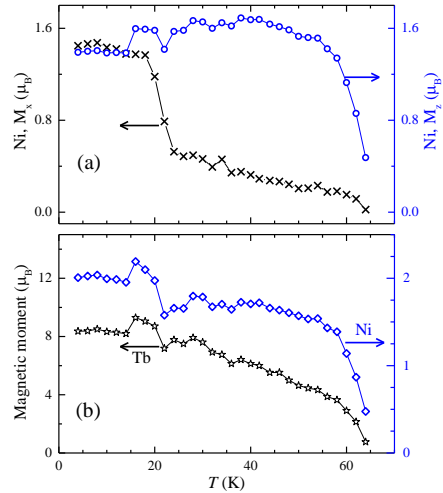

**Fig. S2:** Temperature dependence of (a) *x*- and *z*- components of magnetic moment of Ni and (b) the net magnetic moment of Ni and Tb moment, for $Tb_2BaNiO_5$ obtained from ND data. The lines through the data points serve as guides to the eyes. The error bar in the values is typically 1%. These values are obtained using FullProf program for analyzing ND patterns.

In support of the canting of the magnetic moments discussed in the main text, there is a noticeable change in the values of the Ni/Tb magnetic moments, particularly in the *x*-component ($M_x$) of Ni along *a*-axis, as the temperature is lowered below 25 K, as shown in Fig. S2. The ultimate values at 2 K obtained here are in good agreement with those reported in Ref. 32 in the main text.

**Spontaneous polarization:**
We have calculated the spontaneous polarization (Ps) from the positional coordinates from Rietveld refined coordinates, unit cell parameters and first-principles derived Born effective charges (BEC) taken from the literature [2, 3] using the following relationship:
$P = e/V \sum_k z'_k \Delta(k)$
The Ps can be calculated along the three crystallographic direction, and the net sum of the polarization is expressed by the above equation.

In this equation, the sum runs over all the ions inside the unit cell, while Delta k) is the displacement of the kth ion from its ideal position, zk the Born effective charge for kth ion,
and V the volume of the primitive unit cell [4].

# A gist of the magnetic, magnetoelectric studies on the series $Tb_{2-x}Y_xBaNiO_5$

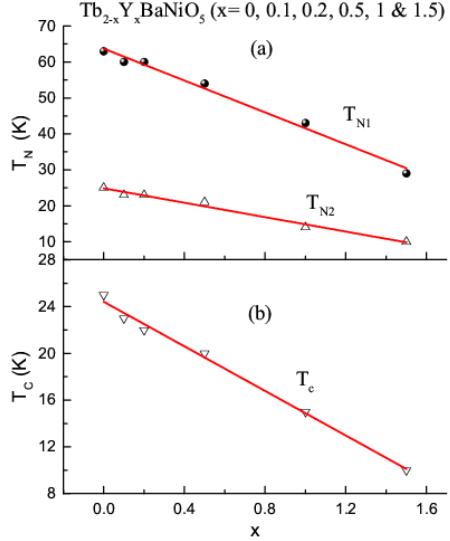

**Fig. S3: (a)** The variation of Néel temperature ($T_{N1}$) and another characteristic temperature ($T_{N2}$), with the dilution of Tb sublattice by (non-magnetic) Y, as determined by magnetic measurements. **(b)** The x-dependence of temperature at which electric dipole ordering sets in, $T_C$ (as inferred from the features in the temperature dependence of dielectric constant and bias current) is plotted.

# Tables following neutron diffraction analysis

**Table I**: Parameters obtained by refining the 80 K, 50 K, 20 and 2 K neutron diffraction patterns of $Tb_2BaNiO_5$ in the orthorhombic *Immm* structure are shown in this Table.

T = 80 K

| Atom | Site | x | y | z | $B_{iso}$ |
|------|------|--------|-------------|-------------|-----------|
| Tb   | 4j   | 0.5000 | 0.0000      | 0.20341(14) | 0.009(42) |
| Ba   | 2c   | 0.5000 | 0.5000      | 0.0000      | 0.366(87) |
| Ni   | 2a   | 0.0000 | 0.0000      | 0.0000      | 0.570(53) |
| O1   | 8l   | 0.0000 | 0.24134(39) | 0.14719(13) | 0.652(41) |
| O2   | 2b   | 0.5000 | 0.0000      | 0.0000      | 0.598(70) |

T = 50 K

| Atom | Site | x | y | z | $B_{iso}$ |
|---|---|---|---|---|---|
| Tb | 4j | 0.5000 | 0.0000 | 0.20250(13) | 0.013(28) |
| Ba | 2c | 0.5000 | 0.5000 | 0.0000 | 0.321(21) |
| Ni | 2a | 0.0000 | 0.0000 | 0.0000 | 0.721(33) |
| O1 | 8l | 0.0000 | 0.23798(26) | 0.14626(10) | 0.637(14) |
| O2 | 2b | 0.5000 | 0.0000 | 0.0000 | 0.597(78) |

T = 20 K

| Atom | Site | x | y | z | $B_{iso}$ |
|---|---|---|---|---|---|
| Tb | 4j | 0.5000 | 0.0000 | 0.20315(16) | 0.023(37) |
| Ba | 2c | 0.5000 | 0.5000 | 0.0000 | 0.311(21) |
| Ni | 2a | 0.0000 | 0.0000 | 0.0000 | 0.607(37) |
| O1 | 8l | 0.0000 | 0.23892(34) | 0.14591(11) | 0.674(34) |
| O2 | 2b | 0.5000 | 0.0000 | 0.0000 | 0.571(23) |

T = 2 K

| Atom | Site | x | y | z | $B_{iso}$ |
|---|---|---|---|---|---|
| Tb | 4j | 0.5000 | 0.0000 | 0.20278(11) | 0.017(63) |
| Ba | 2c | 0.5000 | 0.5000 | 0.0000 | 0.205(10) |
| Ni | 2a | 0.0000 | 0.0000 | 0.0000 | 0.574(41) |
| O1 | 8l | 0.0000 | 0.24083(31) | 0.14733(17) | 0.614(76) |
| O2 | 2b | 0.5000 | 0.0000 | 0.0000 | 0.594(60) |

**DFT calculation methodology and details**

DFT calculations were performed using the Perdew-Burke-Ernzerhof Generalized Gradient Approximation (PBE-GGA) [5] and GGA+U [6] approximation. We also carried out the noncollinear and collinear magnetic calculations within the GGA+U+SO approximation. Structural optimization of lattice parameters and atomic coordinates are performed until the Hellmann–Feynman force on each atom is less than 5 meV/Å. An energy cut-off of 450 eV was used for the plane waves in the basis set while a 2x4x2 Monkhorst-Pack **k**-mesh centered at Γ was used for performing the Brillouin zone integrations. The self-consistent electronic structure calculations, were performed till an energy difference of $10^{-4}$ eV was achieved.

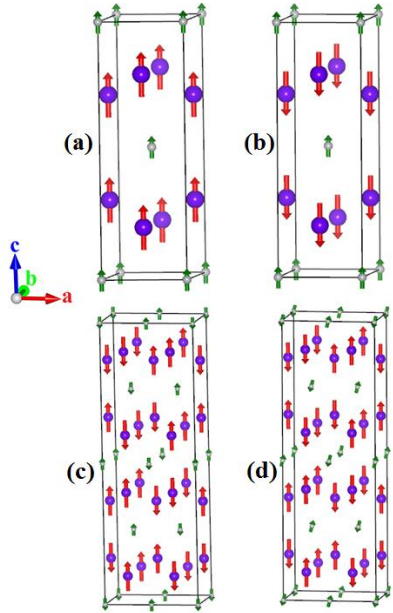

**Fig. S4**: Various magnetic structures of $Tb_2BaNiO_5$ considered in the calculations: a) Ferromagnetic, b) ferrimagnetic, c) collinear anti-ferromagnetic (AFM-CL) and d) Non-collinear antiferromagnetic (AFM-NCL) structures. Tb and Ni ions are shown in violet and grey colours. One crystallographic unit cell is shown in (a) and (b) whereas 2x1x2 supercell is shown in (c) and (d).

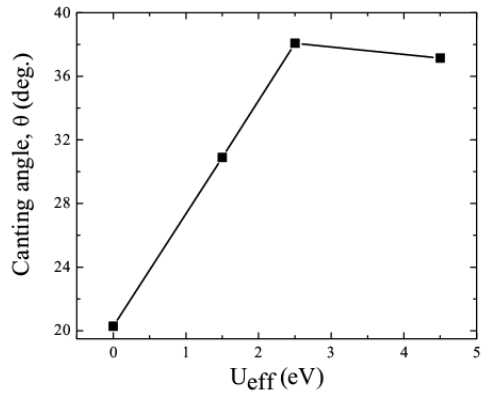

**Fig. S5.** Dependence of canting angle of Ni moments in AFM-NCL magnetic structure on $U_{eff}$ applied to Ni ions, keeping $U_{eff}$ for Tb fixed at 7eV.

**Table II.** Relative energy (in meV) of antiferromagnetic configurations within GGA+U+SO approximation for three different $U_{eff}$ values.

| $U_{eff}$ (eV) | Tb = 0.0, Ni = 0.0 | Tb = 7.0, Ni = 1.5 | Tb = 7.0, Ni = 2.5 |
|---|---|---|---|
| AFM-CL | 13.20 | 152.30 | 0.0 |
| AFM-NCL | 0.0 | 0.0 | 63.60 |

**Table III.** Magnetic moments and corresponding canting angle of $Ni^{2+}$ within GGA+U+SO as a function of $U_{eff}$ applied to Ni. $U_{eff}$ for Tb is kept fixed at 7eV. The values tend towards those observed experimentally from the analysis of the neutron diffraction pattern at 2 K for U>1.5 eV.

| $U_{eff}$ (eV) (Tb, Ni) | $|M_x|(\mu_B)$ | $|M_z|(\mu_B)$ | canting angle(°) |
|---|---|---|---|
| (7.0, 0.0) | 0.48 | 1.29 | 20.3 |
| (7.0, 1.5) | 0.76 | 1.26 | 30.9 |
| (7.0, 2.5) | 0.94 | 1.2 | 38.1 |
| (7.0, 4.5) | 0.98 | 1.3 | 37.2 |

**Table IV.** Experimental and optimized lattice parameters, bond angles and bond lengths for FM and AFM-CL within GGA approximation.

| Lattice constant (Å) | experimental | AFM_calculated | FM_calculated |
|---|---|---|---|
| a | 3.7789 | 3.7807 | 3.7941 |
| b | 5.7780 | 5.8173 | 5.8097 |
| c | 11.3797 | 11.4379 | 11.4309 |
| **Bond angles (degree)** | | | |
| Ni - O2 -Ni | 180.00 | 179.94 | 180.00 |
| O1 - Ni - O1 | 78.89 | 78.59 | 78.60 |
| **Bond lengths (Å)** | | | |
| Ni – O1 | 2.1806 | 2.2064 | 2.2037 |
| Ni – O2 | 1.8894 | 1.8731 | 1.8970 |

**References:**

[1]. Sanjay K Upadhyay and E.V. Sampathkumaran (communicated for publication).
[2]. J. B. Neaton, C. Ederer, U. V. Waghmare, N. A. Spaldin, and K. M. Rabe, Phys. Rev. B 71, 014113 (2005)
[3]. Ph. Ghosez, E.Cockayne, U. V. Waghmare, and K. M. Rabe, Phys. Rev. B 60, 836 (1999).
[4]. Arun Kumar, S. D. Kaushik, V. Siruguri,and Dhananjai Pandey, Phys. Rev B 97, 104402 (2018).
[5]. John P. Perdew, Kieron Burke, and Matthias Ernzerhof, Phys. Rev. Lett.**77**, 3865 (1996).
[6]. V. I. Anisimov, I. V. Solovyev, M. A. Korotin, M. T. Czyzyk, and G. A. Sawatzky, Phys. Rev. B **48**, 16929 (1993).